\documentclass[aps,twocolumn,superscriptaddress]{revtex4-2}
\usepackage{amsmath,amssymb}
\usepackage{graphicx}
\usepackage{booktabs}
\usepackage{hyperref}
\DeclareMathOperator*{\med}{median\,}

\newcommand{\optionalgraphic}[2]{%
  \IfFileExists{#2}{\includegraphics[width=#1]{#2}}{%
    \fbox{\parbox[c][0.115\textheight][c]{\dimexpr#1-2\fboxsep-2\fboxrule\relax}{\centering\scriptsize Figure asset not supplied\par\emph{Image placeholder}}}%
  }%
}
\begin{document}
\title{Unlocking Low-Frequency Laser Doppler Holography Spectra for Functional Retinal and Choroidal Microvascular Phenotyping}
\author{Zacharie Auray}
\author{Adrien Gordon}
\author{Gabriel Cellier}
\author{Yann Fischer}
\author{Michael Atlan}
\affiliation{Langevin Institute, ESPCI Paris, PSL University, CNRS, Paris, France}
\date{\today}
\begin{abstract}
The low-frequency part of optical fluctuation spectra recorded by laser Doppler holography at a frame rate of 37~kHz is typically filtered out to reveal functional blood-flow waveforms in major retinal and choroidal vessels. We construct a time-varying spectral-excess (SE) distribution in capillary-associated territories by selecting complete cardiac cycles and applying successive spatial and beat medians in regions where vessels are unresolved and intermittent eye motion can dominate the signal. At each phase, spectral excess compares the integrated Doppler spectral area between two fixed low-frequency peaks near $\pm2$~kHz with a reference area defined by the high-frequency tail near $\pm18.5$~kHz. The resulting waveform, displayed against normalized cardiac phase, retains variation throughout the cardiac cycle and provides a basis for exploring its temporal organization. We summarize spectral-excess strength by the spectral excess index (SEI), defined as the temporal median of SE, and also describe an alternative evaluation after spectral pooling. A representative recording illustrates the construction. Physiological specificity and repeatability remain to be quantified.
\end{abstract}
\maketitle
\section{Introduction}
Retinal laser Doppler holography (LDH) uses an ultrahigh-speed camera to measure frequency-resolved fluctuations in light scattered by moving blood cells across the eye fundus. At a recording frame rate of 37~kHz, frequencies below 6~kHz are typically filtered out to reveal functional blood-flow waveforms in major retinal and choroidal vessels \cite{Puyo2015,Puyo2018,Puyo2019,Auray2025UltrahighSpeed}. Spectral shape in capillary-associated territories offers a complementary observable, although its interpretation depends on scattering geometry, multiple scattering, hematocrit, vessel density, and motion \cite{Bonner1981,Rajan2009,Fredriksson2013,Carp2023}.

Conventional phase-resolved Doppler OCT measures axial motion through optical phase shifts. In capillaries, small axial phase increments can approach the phase-noise floor, particularly for transverse flow, making the choriocapillaris challenging to assess \cite{Puyo:25}. Speckle variance, decorrelation time, and spectral broadening provide complementary motion sensitivity. With an appropriate optical model, decorrelation times and spectral widths can support velocity estimates in physical units, although the inference is less direct than measuring an axial Doppler shift \cite{Srinivasan2010QuantitativeOCT,Puyo:25}. OCT angiography (OCTA) uses variations among repeated B-scans to form depth-resolved microvascular maps \cite{Makita2006Angiography,Wang2007OpticalAngiography}.

Temporal sampling is critical for recovering capillary fluctuations. Migacz et al. improved choriocapillaris and choroidal OCTA contrast using a 1.6-MHz Fourier-domain mode-locked laser and B-scan rates near 2~kHz \cite{Migacz2019MHzOCTA}, faster than the few-hundred-hertz repetition associated with conventional 3--5~ms interscan intervals \cite{Wei2019HDROCTA}. Puyo et al. used 2-kHz repeated OCT B-scans to resolve heartbeat-induced choriocapillaris spectral broadening and quantify it with a spectral-width index in hertz \cite{Puyo:25}. Their recorded OCT spectrum was concentrated below approximately 1~kHz in that configuration.

A key distinction between these modalities is the dynamic light-scattering regime selected by optical sectioning. OCT and OCTA use coherence gating to localize the signal in depth and reduce multiply scattered contributions, whereas conventional en face LDH and laser speckle contrast imaging (LSCI) lack this depth gate \cite{Srinivasan2010QuantitativeOCT,Puyo:25}. At a given en face retinal image location, photons detected without depth gating may have accumulated Doppler shifts through successive interactions with moving blood cells. This is expected to favor broader fluctuation spectra and faster decorrelation for comparable velocities \cite{Bonner1981,Carp2011}.

These findings motivate analysis of capillary-associated spectral dynamics in LDH. Near 2~kHz, representative fundus images show diffuse contrast with unresolved vessel contours. Intermittent saccades produce broadband streaks that obscure weaker, recurrent spectral features \cite{Puyo2018,Puyo2019,Puyo2020}. We seek to recover these features without optical sectioning by selecting regions without visible retinal or choroidal vessels, retaining complete cardiac cycles, and applying spatial medians followed by aggregation at corresponding phases across beats. Spatial selection limits contributions from resolved vessels, while median aggregation reduces intermittent motion artifacts. Together, these steps reveal low-frequency spectral structure and its variation throughout the cardiac cycle.

\subsection{OCTA and the retinal and choroidal microvascular beds}
The retinal and choroidal circulations support different neural territories. Radial peripapillary capillaries accompany nerve-fiber-layer axons; the macular superficial, intermediate, and deep plexuses supply the inner retina. Projection-resolved OCTA has clarified these layered networks and reduced projection artifacts.\cite{Henkind1967RPC,Yu2015RPC,Campbell2017RetinalPlexuses} The choriocapillaris supports the retinal pigment epithelium (RPE) and photoreceptors beneath Bruch's membrane.\cite{McLeod1994Choroid} Table~\ref{tab:capillary_beds} summarizes these biological contexts.

OCTA enables assessment of capillary density, nonperfusion, foveal avascular zone geometry, and choriocapillaris flow-signal deficits.\cite{Campbell2017RetinalPlexuses,Spaide2016,Sun2019DRProgression} Density and flow-void metrics describe the distribution of detectable flow without directly measuring absolute perfusion.\cite{Zhang2018CCFlowVoids} Sparse repeats and interscan timing constrain spectral characterization and determine sensitivity and decorrelation saturation.\cite{Wei2019HDROCTA} Segmentation, attenuation, projection artifacts, and image quality also affect the visible network. En face LDH integrates scattering over depth, so its spectral signal cannot presently be assigned to an individual plexus.

\subsection{Glaucoma: peripapillary and macular microvasculature}
OCTA has demonstrated reduced optic-disc perfusion and vascular density in glaucoma.\cite{Jia2014GlaucomaOCTA} Lower baseline macular and optic-nerve-head vessel density has been associated with faster retinal nerve-fiber-layer (RNFL) thinning,\cite{Moghimi2018GlaucomaVD} and faster initial optic-nerve-head capillary-density loss with later visual-field progression.\cite{Tansuebchueasai2024ONH} Vascular and neural changes are coupled, so these associations do not establish vascular causality. RNFL and ganglion-cell structure, visual fields, and macular and optic-disc OCTA provide relevant comparators for functional spectral measurements. Optic-nerve-head pulsatility measured by laser speckle flowgraphy supplies a related dynamic comparator.\cite{Enomoto2021}

\subsection{Diabetic retinopathy: plexus-specific ischemia and vascular function}
Diabetic retinopathy (DR) affects macular plexuses unevenly. Prospective OCTA studies have linked deep-plexus foveal avascular zone geometry and vessel density to DR progression, and superficial-plexus density to edema development.\cite{Sun2019DRProgression} More recent studies extend prediction to visual decline and deep-capillary-plexus perfusion deficits.\cite{Yang2026DRPrediction,Zhuang2026DCP} Parapapillary choroidal measurements have also been investigated as longitudinal predictors.\cite{Guo2023Parapapillary} Impaired flicker-evoked retinal vasodilation provides a functional comparator.\cite{Mandecka2007Flicker} These findings motivate assessment of cardiac spectral dynamics alongside anatomy, disease stage, metabolic status, treatment, and optical quality.

\subsection{AMD: choriocapillaris support and early progression}
In age-related macular degeneration (AMD), the choriocapillaris--Bruch's membrane--RPE interface supports photoreceptor metabolism.\cite{McLeod1994Choroid} OCTA flow-signal deficits have been associated with progression,\cite{Corvi2021,Pu2026ALSTAR2} and with visual dysfunction, especially delayed rod-mediated dark adaptation.\cite{Kar2023ChoriocapillarisRMDA} Drusen, RPE alterations, attenuation, and segmentation affect the measured signal, while retinal and deeper choroidal scattering contribute to LDH. Structural OCT, choriocapillaris OCTA, and dark adaptation are therefore relevant comparators for investigating a functional spectral readout.

\begin{table*}[t]
\caption{Microvascular territories and established comparators motivating the spectral-excess analysis. These are biological contexts, not proven LDH depth assignments.}
\label{tab:capillary_beds}
\begin{ruledtabular}
\begin{tabular}{p{0.23\textwidth}p{0.31\textwidth}p{0.35\textwidth}}
Territory & Biological and clinical context & Established comparators \\
\hline
Radial peripapillary capillaries & Support of nerve-fiber-layer axons; glaucoma & RNFL and ganglion-cell structure, visual field, optic-nerve-head and macular OCTA vessel density \\
Macular retinal plexuses & Inner-retinal support; DR and macular ischemia & Plexus-resolved OCTA, nonperfusion, foveal avascular zone geometry, DR stage and edema \\
Choriocapillaris & RPE and photoreceptor support; AMD & OCTA flow-signal deficits, drusen and RPE structure, rod-mediated dark adaptation \\
\end{tabular}
\end{ruledtabular}
\end{table*}

\subsection{Position and objectives}
Doppler OCT, laser speckle methods, and ocular ultrasound provide complementary hemodynamic measurements with different spatial, temporal, and depth sensitivities.\cite{Yazdanfar2003,Choi2012,BoasDunn2010,Urs2016} LDH combines full-field spectra with cardiac waveform measurements.\cite{Puyo2019Choroidal,Giammanco2026} Vessel-excluded regions can contain unresolved microvessels, larger-vessel leakage, and nonvascular scattering, so the signal is capillary-associated without being anatomically specific.\cite{RivaPetrig1995,Puyo:25}

The main result is the construction of the spectral-excess waveform $\mathrm{SE}(t)$ over an aligned cardiac cycle, displayed against normalized phase $t/T$. SEI summarizes its strength as a temporal median; an alternative pooled-spectrum evaluation is also defined. A representative recording illustrates the construction. The retained temporal information provides a basis for subsequent exploration of waveform organization, including descriptors discussed by Buisson et al.\cite{Buisson2026WaveformShape}

\section{Experimental methods}
\label{sec:experimental_methods}
The optical system and recording conditions follow the related LDH source-study protocol.

\subsection{Optical system}

\begin{figure}[t]
\centering
\optionalgraphic{0.98\linewidth}{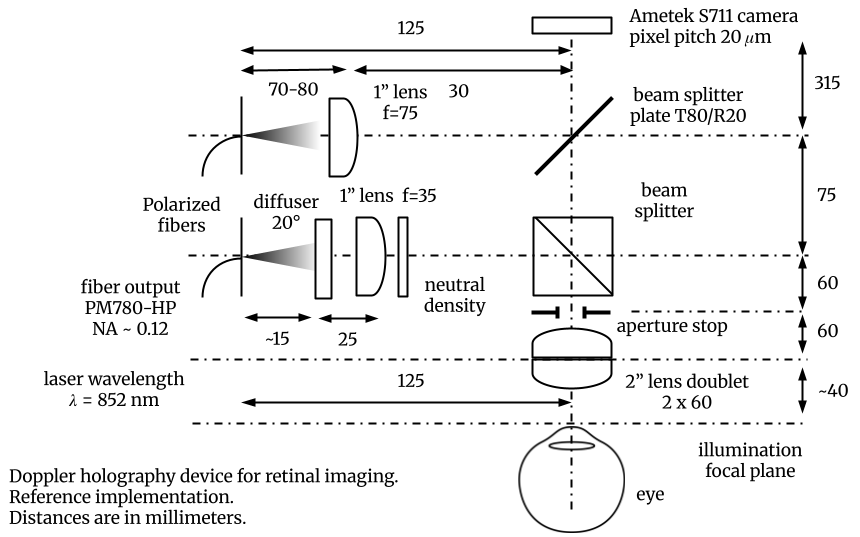}
\caption{Optical layout of the 852-nm on-axis laser Doppler holography ophthalmoscope. Fiber-delivered light is divided between illumination and reference paths. The diffuser and relay optics provide retinal illumination; fundus backscatter is recombined with the reference field and recorded interferometrically with the high-speed camera. The diagram shows the optical arrangement used to obtain the holograms from which full-field Doppler power spectra are reconstructed.}
\label{fig:setup}
\end{figure}

The clinical prototype and optical configuration used for ultrahigh-speed laser Doppler holography are shown in Fig.~\ref{fig:setup}.

Retinal interferograms were acquired with an on-axis digital holography ophthalmoscope based on an inline Mach--Zehnder interferometer operating at $\lambda=852$~nm.\cite{Auray2025UltrahighSpeed} Light from a diode laser (Thorlabs FPV852P) was divided between the illumination and reference arms with a $90/10$ power ratio using polarization-maintaining fibers. In the illumination arm, an engineered diffuser (Thorlabs ED1-C20-MD, $20^\circ$ circular top-hat) and two 60-mm biconvex lenses provided diffuse Maxwellian illumination.\cite{Bratasz2022Diffuse} Cross-polarized light backscattered by the fundus was recombined with the reference beam and recorded with a Phantom S711 camera over $512\times320$ pixels with a $20~\mu\mathrm{m}$ pixel pitch.

\subsection{Recording conditions}

Retinal holograms for the representative recording were acquired without pharmacological pupil dilation in one eye of a healthy 25-year-old man after written informed consent. During acquisition, the participant viewed a static fixation target with the contralateral eye. The procedures adhered to the Declaration of Helsinki and were authorized by the local ethics boards. The optical power at the diffused focal spot entering the eye was 5.0~mW. In the reported assessment of this 852-nm diffuse Maxwellian configuration, the most conservative design recommendation was an operating power not exceeding approximately 6~mW, set by the ISO 15004-2:2024 iris estimate; the ANSI Z80.36-2021 retinal design estimate was approximately 11~mW, and the anterior-segment condition was not limiting.\cite{Bratasz2022Diffuse}

The acquisition contained 163,840 interferograms recorded over approximately 4.4~s at a frame rate of 37~kHz; four complete cardiac cycles were retained for analysis.\cite{Auray2025UltrahighSpeed} Acquisition and live visualization can be supported by Holovibes, a GPU-accelerated framework for real-time hologram reconstruction, high-rate camera streaming and concurrent recording.\cite{Dubosc2025Holovibes} The spectral-excess waveform and index are derived in the subsequent analysis.

\section{Methods}
\subsection{Reconstruction and temporal demodulation of the Doppler spectrum}
The recorded interferograms were numerically propagated from the camera plane to the retinal conjugate plane by a Fresnel transform at a reconstruction distance of 480~mm. The transform combines a quadratic spatial phase factor with a two-dimensional Fourier transform to recover the complex retinal field $H(x,y,n)$ at camera frame $n$.\cite{Puyo2018,Auray2025UltrahighSpeed} In each non-overlapping block of 256 frames, the complex field was arranged as a space--time matrix and decomposed by singular-value decomposition (SVD). The 24 components associated with the largest singular values were removed before reconstruction of the filtered field, to suppress dominant spatially coherent clutter and bulk eye motion.\cite{Puyo2020} A 256-point temporal fast Fourier transform (FFT) of the filtered complex field at each pixel then yielded the complex Doppler amplitude $E(x,y,\tau,\omega)$ and its power spectrum $S\sim\left|E\right|^2$, where $\tau$ indexes the acquisition-time window and $\omega$ is Doppler angular frequency. At $f_s=37$~kHz, each window spans 6.92~ms and the native FFT-bin spacing is 144.5~Hz. Averaging the power in each group of four adjacent native FFT bins retains the full signed spectrum as 64 exported frequency bins, with calibrated spacing $\delta f=4f_s/256=0.578125$~kHz. EyeFlow v1.24.0 assigns each exported bin the mean signed frequency of its four native bins. Frequencies are expressed in kilohertz as $f=\omega/(2\pi)$, with the positive and negative branches kept in their stored FFT order; the Nyquist limit is $\pm18.5$~kHz.\cite{EyeFlow2026Gap} The time-stamped spectra are subsequently registered to a common normalized cardiac-phase grid $t/T$ and complete beat $b$. The SVD rejection rank is a fixed processing choice; residual motion is addressed by subsequent robust pooling rather than assumed to be completely removed at this stage.

\subsection{Recovering low-frequency spectra by spatial filtering and complete-beat pooling}
\begin{figure}[t]
\centering
\optionalgraphic{0.48\linewidth}{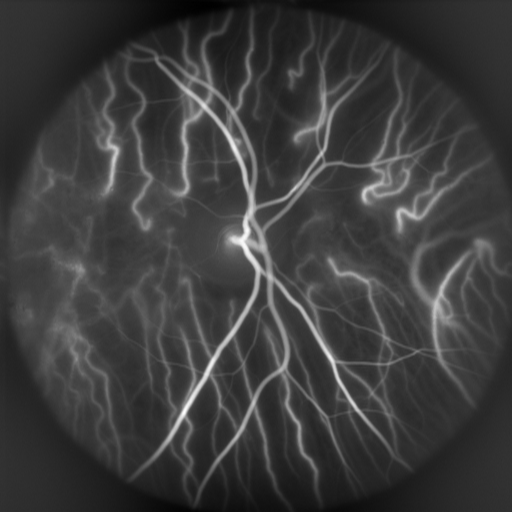}\hfill
\optionalgraphic{0.48\linewidth}{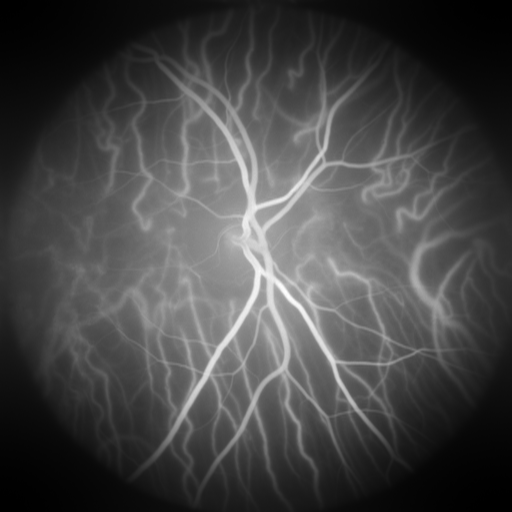}\\[-0.2em]
\makebox[0.48\linewidth]{\small (a)}\hfill
\makebox[0.48\linewidth]{\small (b)}\par\smallskip
\optionalgraphic{0.48\linewidth}{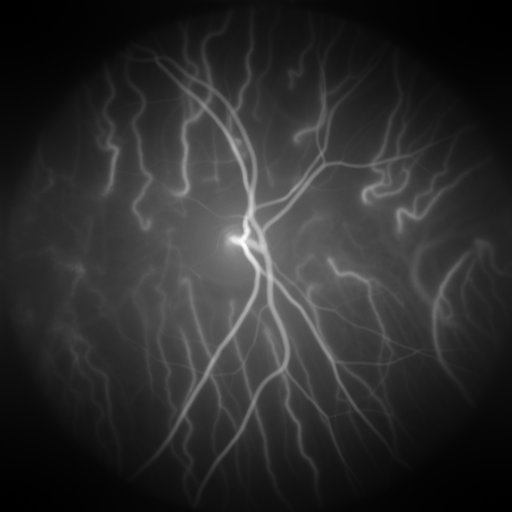}\hfill
\optionalgraphic{0.48\linewidth}{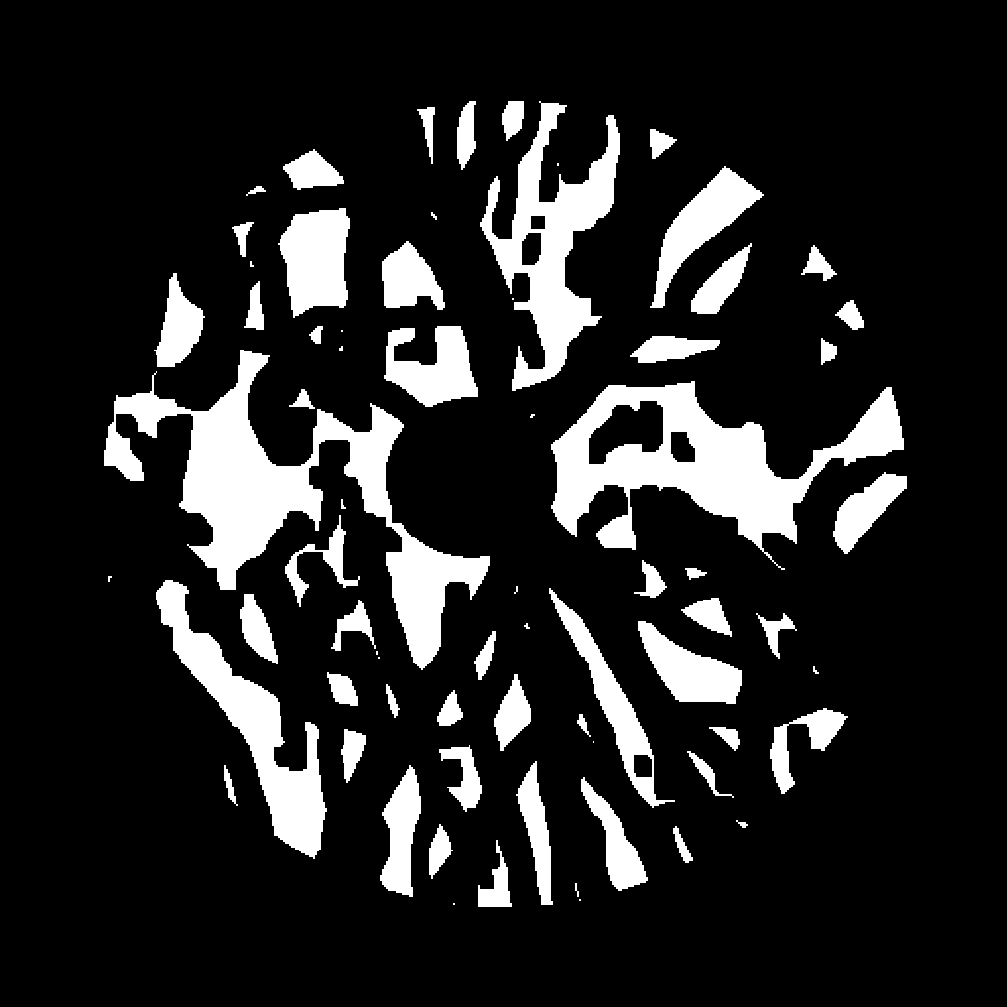}\\[-0.2em]
\makebox[0.48\linewidth]{\small (c)}\hfill
\makebox[0.48\linewidth]{\small (d)}
\caption{Illustrative fundus maps and spatial support for the spectral analysis. (a) Flat-field-corrected zeroth Doppler moment; (b) 3--9~kHz band-integrated Doppler power; (c) 9--18~kHz Doppler power; and (d) the vessel-exclusion mask. In (d), white denotes retained pixels and black denotes excluded vessels, optic-disc and outer-field regions. The same retained spatial support is applied to every exported spectral bin.}
\label{fig:images}
\end{figure}

\begin{figure*}[t]
\centering
\optionalgraphic{0.78\textwidth}{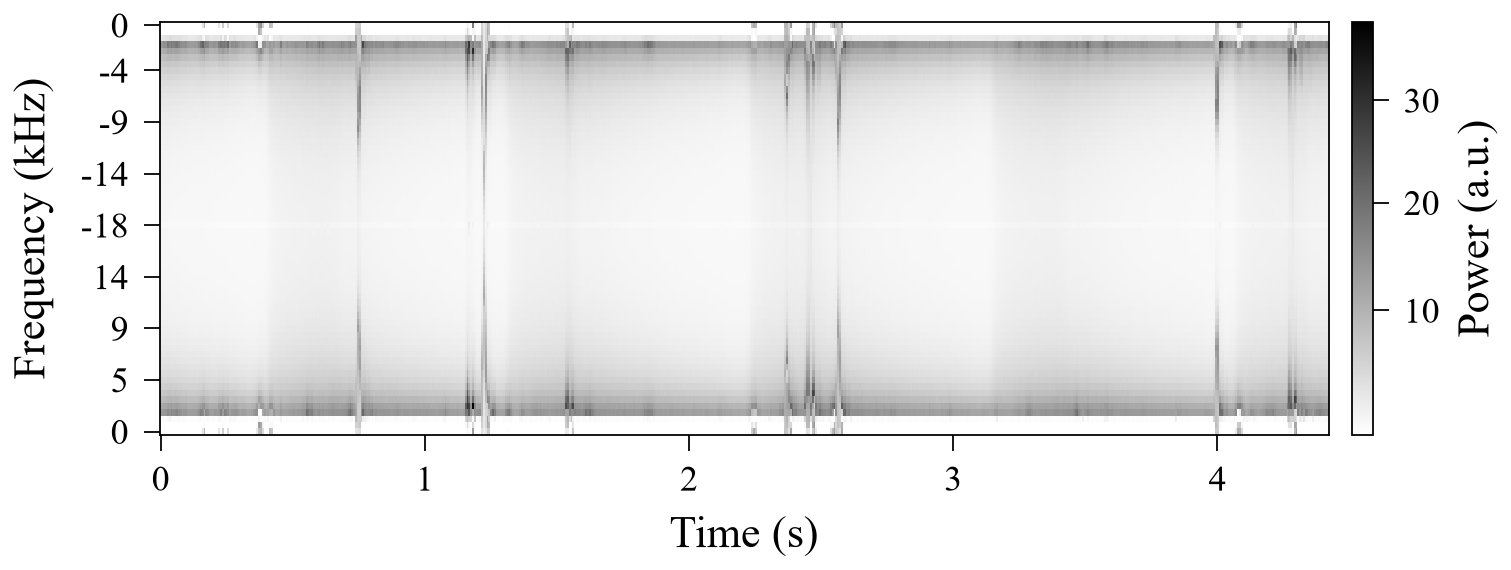}
\caption{Acquisition-time Doppler spectrogram from the vessel-excluded retinal region over an approximately 4.4-s recording, before cardiac-phase registration and aggregation of the four retained complete beats. Narrow, broadband transient streaks interrupt the otherwise more persistent frequency structure and are consistent with brief eye-motion events. Such nonrecurrent excursions motivate the spatial median and subsequent cardiac aggregation used to construct $S(\omega)$ and $S(t,\omega)$. The physical-frequency labels are in kilohertz; the positive and negative branches retain their stored FFT order.}
\label{fig:acquisition_spectrogram}
\end{figure*}

\begin{figure*}[t]
\centering
\optionalgraphic{0.78\textwidth}{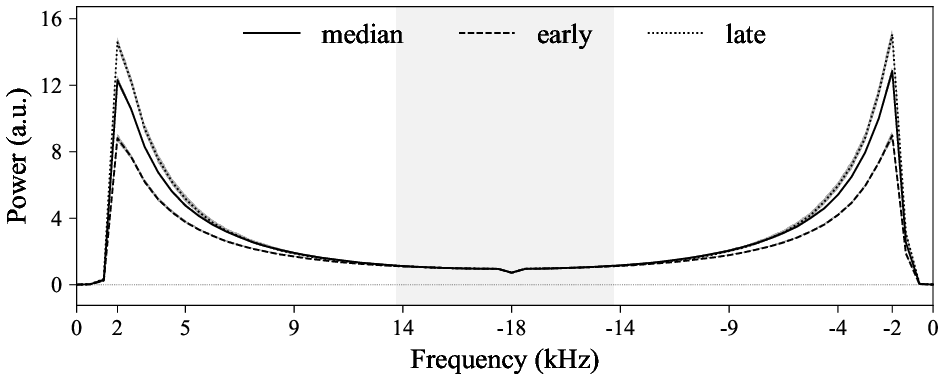}
\caption{Spectra underlying the spectral-excess construction. Solid, dashed, and dotted curves show the pooled spectrum and its early-phase (first third) and late-phase (remaining two thirds) illustrations. Both signed-frequency blocks retain their stored order. Each curve is divided by its own central-valley median for display; this display normalization is not applied before calculating the spectral areas. The pale envelopes reproduce the dispersion in the source plot and the gray background highlights the valley region. The peak indices are the fixed bins nearest $\pm2$~kHz. The pooled spectrum supplies the pooled-spectrum index; the phase-resolved spectrum supplies $\mathrm{SE}(t)$. Physical frequencies are displayed in kilohertz in the stored positive-then-negative FFT order.}
\label{fig:aggregated_spectrum}
\end{figure*}

Ultrahigh-speed LDH yields Doppler power spectra over the fundus. At low frequency, spatial patterns are predominantly diffuse rather than vessel-resolved, so spectral shape---not vessel morphology---is the analysis target. Recovery combines three complementary operations. First, spatial discrimination and filtering exclude segmented visible retinal and choroidal vessels and the optic disc within a central field, using independently obtained vascular masks rather than low-frequency vessel contrast. Related Doppler-hologram artery--vein segmentation has been shown to benefit from cardiac features,\cite{Dubosc2025RetinalSegmentation} although this does not establish capillary specificity for the retained region. Retained-mask coverage is recorded. Second, cardiac registration retains four complete beats and discards incomplete edge cycles. Phase zero is the maximum time derivative of the arterial waveform, corresponding to maximal systolic acceleration. Each cycle extends to the next occurrence of this landmark and is rescaled to normalized phase $t/T$. Third, the median spectrum is calculated first over retained vessel-excluded pixels at each acquisition time and exported spectral bin; finite negative spatial-median values are truncated to zero. These spatial medians are then linearly interpolated within each complete beat onto a common cardiac-phase grid. Let $S(t,b,\omega)$ denote these phase-aligned spatial-median spectra. EyeFlow v1.24.0 forms the pooled spectrum by taking one median over all valid phase--beat samples at each frequency:
\begin{equation}
S(\omega)\equiv\med_{t,b}\!S(t,b,\omega),
\label{eq:Srobust}
\end{equation}
while the cardiac-phase-resolved spectrum is
\begin{equation}
S(t,\omega)\equiv\med_b\!S(t,b,\omega).
\label{eq:Stbw}
\end{equation}
Here $t$ is the time coordinate within the registered cardiac cycle, $t/T$ is its normalized phase, and $b$ indexes complete beats; $\tau$ remains the acquisition-time coordinate. Each beat is resampled on the same phase grid. In Eq.~\eqref{eq:Srobust}, $\med_{t,b}$ denotes a single joint median of the phase--beat samples, while Eq.~\eqref{eq:Stbw} takes a median across beats separately at each phase. Consequently, $S(\omega)$ need not equal $\med_t S(t,\omega)$: the latter is a nested median, and medians are not generally associative. Both reductions follow the spatial median, omit nonfinite phase--beat samples, and retain the input bin order. This spatial reduction followed by cardiac aggregation can attenuate isolated, nonrecurrent saccadic streaks. Figure~\ref{fig:images} illustrates the selection of measurement regions, while Figs.~\ref{fig:acquisition_spectrogram} and \ref{fig:aggregated_spectrum} illustrate motion-related streaks in acquisition-time spectra and their compact spectral representation.

EyeFlow v1.24.0 supplies the reference code for mask alignment, spatial reduction, cardiac registration, joint phase--beat spectral pooling, and the spectral-area calculation.\cite{EyeFlow2026Gap} The plotted physical frequencies are retained throughout this analysis. Areas are evaluated in the unwrapped stored-bin coordinate with the calibrated positive bin spacing $\delta f$, rather than integrating across the discontinuity in the signed display axis. A common change in this spacing rescales $A$ and $A_0$ equally and cancels in their ratio.

\subsection{Construction of the spectral-excess distribution}
The two peak indices, $j_{\mathrm L}$ and $j_{\mathrm R}$, are selected as the exported bins nearest $+2$ and $-2$~kHz, respectively, and remain fixed throughout the cardiac cycle. With the above calibration, their centers are approximately $+1.95$ and $-2.10$~kHz. At each phase, $A(t)$ is the trapezoidal area under $S(t,\omega)$ between these indices in stored-bin order. The reference height $h_{\mathrm v}(t)$ is the median power in a fixed neighborhood around the central high-frequency valley. The reference implementation uses the central quarter of the stored bins, comprising 16 bins for this 64-bin spectrum, giving the reference rectangle $A_0(t)=(j_{\mathrm R}-j_{\mathrm L})\delta f\,h_{\mathrm v}(t)$. Spectral excess is
\begin{equation}
\mathrm{SE}(t)=\frac{A(t)-A_0(t)}{A_0(t)}
=\frac{A(t)}{A_0(t)}-1.
\label{eq:shape}
\end{equation}
Computing this ratio at every registered phase constructs the $\mathrm{SE}(t)$ distribution over the cardiac cycle. The peak indices remain fixed, while the central-valley reference is recomputed at each phase. Integration follows the stored signed-bin order; the common frequency-spacing factor $\delta f$ cancels in the ratio.

\begin{figure}[t]
\centering
\optionalgraphic{0.99\linewidth}{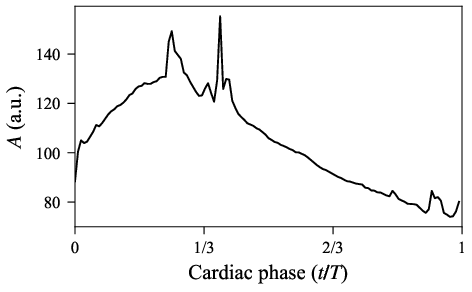}\par\smallskip
\optionalgraphic{0.99\linewidth}{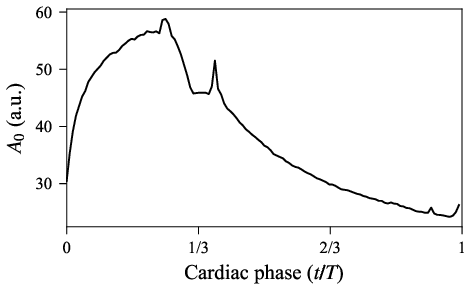}\par\smallskip
\optionalgraphic{0.99\linewidth}{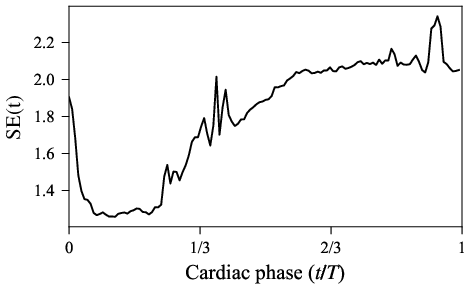}
\caption{Construction of $\mathrm{SE}(t)$ from four complete beats in the representative recording, displayed against normalized cardiac phase $t/T$. Phase zero is the maximum derivative of the arterial waveform (maximal systolic acceleration). From top to bottom: (a) integrated peak-to-peak area $A(t)$; (b) central-valley reference rectangle $A_0(t)$; and (c) spectral excess $\mathrm{SE}(t)=A(t)/A_0(t)-1$. Fixed peak indices and a phasewise central-valley median are used throughout. Comparing the numerator and denominator traces helps identify their separate contributions to the waveform. The temporal median of this waveform defines SEI in Eq.~\eqref{eq:sei}.}
\label{fig:SE_t}
\end{figure}

Spectral excess measures valley-normalized relative area. A positive value means that the integrated area exceeds the reference rectangle; zero means equal areas and does not require a perfectly flat spectrum. An increase in $A(t)$ or a decrease in $A_0(t)$ can raise $\mathrm{SE}(t)$, so both area traces are retained alongside the waveform (Fig.~\ref{fig:SE_t}). A common multiplicative spectral gain at a given phase cancels, but additive or frequency-dependent backgrounds do not. A positive, sufficiently stable $A_0(t)$ is required; invalid or insufficient-signal denominators are flagged using prespecified quality criteria. The reference processing uses a NaN-aware central-bin median.\cite{EyeFlow2026Gap}

\subsection{Spectral excess index}
The spectral excess index is
\begin{equation}
\mathrm{SEI}=\med_{t}\!\mathrm{SE}(t).
\label{eq:sei}
\end{equation}
The median is taken over the equally spaced, valid phase samples of the registered cardiac cycle and summarizes the typical spectral-excess strength.

The SEI can also be evaluated after spectral pooling. The pooled-spectrum index differs from a temporal median of $\mathrm{SE}(t)$ (Eq.~\eqref{eq:sei}). For this alternative evaluation, let $A$ be the trapezoidal area of $S(\omega)$ between the fixed peak indices, $h_{\mathrm v}$ the median power in the same central reference neighborhood, and $A_0=(j_{\mathrm R}-j_{\mathrm L})\delta f\,h_{\mathrm v}$ the corresponding reference rectangle. The index is then $(A-A_0)/A_0$. Pooling the spectra before calculating this ratio need not give the median of the phasewise ratios. We denote this alternative by $\mathrm{SEI}_{\mathrm{pool}}=(A-A_0)/A_0$ to distinguish it from Eq.~\eqref{eq:sei}. EyeFlow v1.24.0 exports SEI as the temporal median of valid phasewise $\mathrm{SE}(t)$ values; the pooled-spectrum alternative is defined here as a separate descriptive evaluation.\cite{EyeFlow2026Gap}

Both evaluations are dimensionless measures of relative spectral excess. They do not measure absolute blood flow, blood-cell velocity, or capillary density. Positive, sufficiently stable reference areas are required, with invalid or insufficient-signal phases flagged according to the same quality criteria as the waveform construction.

\subsection{Quality control}
The proposed fundus-to-non-signal-background power ratio near 3~kHz is used to establish sufficient optical signal, not to normalize the spectral-excess calculation. Quality review also considers mask coverage, unstable valley levels, and residual saccadic streaks. Robust aggregation can attenuate isolated artifacts but does not remove sustained contamination.\cite{Kernick1999,Fredriksson2013,Carp2011,Puyo2020}

\section{Results}
\subsection{Recovery of a phase-resolved spectral-excess distribution}
The acquisition-time spectrogram shows intermittent broadband streaks superimposed on persistent spectral structure (Fig.~\ref{fig:acquisition_spectrogram}). Spatial exclusion and aggregation of four complete beats yield spectra with a central valley and peaks in the two signed-frequency blocks (Fig.~\ref{fig:aggregated_spectrum}). Applying the phasewise area ratio produces the $\mathrm{SE}(t)$ distribution shown in Fig.~\ref{fig:SE_t}, with phase zero at maximal arterial systolic acceleration. In this example, spectral excess is lower over part of the early cycle and increases through much of the remaining cycle. The separate $A(t)$ and $A_0(t)$ traces show why this temporal behavior cannot be attributed to the integrated area alone. The recovered waveform is the principal output of the construction: it retains both spectral-excess level and cardiac-phase organization.

\section{Discussion}
The main contribution is the construction of a cardiac-phase-resolved spectral-excess distribution from spatially diffuse, vessel-excluded retinal Doppler spectra. Spatial discrimination, complete-beat selection, and robust aggregation recover a compact waveform even when individual vessels cannot be distinguished in the motivating low-frequency images. The resulting $\mathrm{SE}(t)$ retains the timing and organization of relative spectral excess, while SEI summarizes spectral-excess strength by its median over the cycle.

Both readouts describe a composite optical signal. Retained vascular-gap regions may contain capillaries, unresolved arterioles and venules, residual contributions from visible vessels, and retinal or choroidal scattering. Their relative weights depend on vessel geometry, depth, hematocrit, multiple scattering, and optical sampling. The method does not separate retinal and choroidal contributions, identify a specific vascular plexus, or convert spectral excess into absolute perfusion or velocity.\cite{Rajan2009,Fredriksson2013,Carp2023,Puyo:25} Additive backgrounds, a small valley-reference area, imperfect registration, and residual eye motion can affect both the waveform and its scalar summary. Inspecting $A(t)$ and $A_0(t)$ alongside $\mathrm{SE}(t)$ is therefore an integral part of interpretation.

\subsection{Relationship to OCTA and functional microvascular phenotyping}
OCTA and spectral excess interrogate different aspects of the vascular signal. OCTA localizes detectable flow within retinal and choroidal slabs and quantifies its spatial distribution; $\mathrm{SE}(t)$ describes the cardiac organization of a depth-integrated Doppler spectral shape. This provides a rationale for testing complementarity, not a basis for assuming greater sensitivity or specificity. In glaucoma, relevant comparisons include RNFL and ganglion-cell structure, visual field, and optic-nerve-head or macular OCTA. In DR, they include plexus-specific nonperfusion, edema, and clinical stage. In AMD, they include choriocapillaris flow-signal deficits, drusen/RPE changes, and dark adaptation.\cite{Moghimi2018GlaucomaVD,Tansuebchueasai2024ONH,Sun2019DRProgression,Kar2023ChoriocapillarisRMDA,Pu2026ALSTAR2} Such comparisons are needed to establish physiological specificity and clinical relevance.

The broader interest in functional microvascular phenotyping also extends to systemic disease. Retinal OCTA measurements have been associated with cardiometabolic factors and coronary disease,\cite{Arnould2018EYEMI,Sun2021CardiometabolicOCTA,Yang2025CoronaryOCTA} while retinal vascular reactivity has been investigated in relation to later cardiovascular events.\cite{Theuerle2021RetinalFunctionMACE} These findings motivate future ocular and systemic comparisons of spectral excess. They do not establish that SEI or $\mathrm{SE}(t)$ measures systemic microvascular health. Added information must be evaluated beyond anatomy, OCT/OCTA, image quality, and systemic covariates in independently designed studies.

\subsection{Possibilities for exploring temporal organization}
The full $\mathrm{SE}(t)$ distribution provides a basis for many subsequent descriptions of temporal organization beyond a single strength value. Candidate families include normalized timing and temporal moments, crest width, excursion and persistence, cumulative-shape quantiles, temporal concentration or effective support, normalized slopes, and harmonic organization. The waveform-shape framework of Buisson et al. provides explicit examples of these families for retinal arterial velocity waveforms.\cite{Buisson2026WaveformShape} Adapting such mathematical descriptors to spectral excess would allow signals with similar SEI values to be compared by when, how broadly, and how regularly their excess is distributed through the cycle.

This adaptation would require explicit preprocessing and validity rules, including phase registration, sampling, and any harmonic filtering. Descriptors that treat waveform amplitudes as nonnegative weights would require finite, nonnegative spectral excess and a positive total. Their mathematical use would not transfer the physiological interpretation of arterial velocity or cumulative displacement to $\mathrm{SE}(t)$. No such expanded descriptor panel is calculated or clinically validated here; the present result is the construction of the distribution and its SEI summary.

\section{Conclusion}
Vessel exclusion, complete-beat selection, and spatial reduction followed by cardiac aggregation allow a phase-resolved spectral-excess waveform $\mathrm{SE}(t)$ to be constructed from retinal Doppler spectra. SEI summarizes its strength as the temporal median; an alternative pooled-spectrum evaluation is defined separately. A representative recording illustrates the construction, while the full waveform provides a basis for subsequent exploration of temporal organization. Technical repeatability, physiological specificity, and clinical value remain to be established.

\section*{Open-source code availability}
EyeFlow v1.24.0 supplies the reference vessel-exclusion, cardiac aggregation, and calibrated spectral-processing workflow at \url{https://github.com/DigitalHolography/EyeFlow/releases/tag/v1.24.0}.\cite{EyeFlow2026Gap} Equations~\eqref{eq:shape} and \eqref{eq:sei} define the phase-resolved spectral-excess waveform and its temporal-median index.

\bibliographystyle{apsrev4-2}
\bibliography{doppler_spectral_endpoints}

@article{Bonner1981,
  author = {Bonner, Robert and Nossal, Richard},
  title = {Model for laser Doppler measurements of blood flow in tissue},
  journal = {Applied Optics},
  year = {1981},
  volume = {20},
  pages = {2097--2107},
  doi = {10.1364/AO.20.002097},
  pmid = {20332893}
}

@inproceedings{Puyo2015,
  author = {Puyo, L{\'e}o and Ferezou, Isabelle and Rancillac, Armelle and Simonutti, Matteo and Paques, Michel and Sahel, Jos{\'e}-Alain and Fink, Mathias and Atlan, Michael},
  title = {Pulsatile microvascular blood flow imaging by short-time Fourier transform analysis of ultrafast laser holographic interferometry},
  booktitle = {2015 8th Biomedical Engineering International Conference (BMEiCON)},
  year = {2015},
  doi = {10.1109/BMEiCON.2015.7399525},
  eprint = {1510.01892},
  archivePrefix = {arXiv}
}

@article{Puyo2018,
  author = {Puyo, L{\'e}o and Paques, Michel and Fink, Mathias and Sahel, Jos{\'e}-Alain and Atlan, Michael},
  title = {In vivo laser Doppler holography of the human retina},
  journal = {Biomedical Optics Express},
  year = {2018},
  volume = {9},
  pages = {4113--4129},
  doi = {10.1364/BOE.9.004113},
  pmid = {30615709}
}

@article{Puyo2019,
  author = {Puyo, L{\'e}o and Paques, Michel and Fink, Mathias and Sahel, Jos{\'e}-Alain and Atlan, Michael},
  title = {Waveform analysis of human retinal and choroidal blood flow with laser Doppler holography},
  journal = {Biomedical Optics Express},
  year = {2019},
  volume = {10},
  pages = {4942--4963},
  doi = {10.1364/BOE.10.004942},
  pmid = {31646021}
}

@article{Puyo2020,
  author = {Puyo, L{\'e}o and Paques, Michel and Atlan, Michael},
  title = {Spatio-temporal filtering in laser Doppler holography for retinal blood flow imaging},
  journal = {Biomedical Optics Express},
  year = {2020},
  volume = {11},
  pages = {3274--3287},
  doi = {10.1364/BOE.392699},
  pmid = {32637254}
}

@article{Rajan2009,
  author = {Rajan, Vinayakrishnan and Varghese, Babu and van Leeuwen, Ton G. and Steenbergen, Wiendelt},
  title = {Review of methodological developments in laser Doppler flowmetry},
  journal = {Lasers in Medical Science},
  year = {2009},
  volume = {24},
  number = {2},
  pages = {269--283},
  doi = {10.1007/s10103-007-0524-0},
  pmid = {18236103}
}

@article{Kernick1999,
  author = {Kernick, D. P. and Tooke, J. E. and Shore, A. C.},
  title = {The biological zero signal in laser Doppler fluximetry---origins and practical implications},
  journal = {Pfl\"ugers Archiv},
  year = {1999},
  volume = {437},
  pages = {624--631},
  doi = {10.1007/s004240050826},
  pmid = {10089577}
}

@article{Fredriksson2013,
  author = {Fredriksson, Ingemar and Burdakov, Oleg and Larsson, Marcus and Str\"omberg, Tomas},
  title = {Inverse Monte Carlo in a multilayered tissue model: merging diffuse reflectance spectroscopy and laser Doppler flowmetry},
  journal = {Journal of Biomedical Optics},
  year = {2013},
  volume = {18},
  pages = {127004},
  doi = {10.1117/1.JBO.18.12.127004},
  pmid = {24352692}
}

@article{Carp2011,
  author = {Carp, Stefan A. and Roche-Labarbe, Nad\`ege and Franceschini, Maria-Angela and Srinivasan, Vivek J. and Sakad\v{z}i\'{c}, Sava and Boas, David A.},
  title = {Due to intravascular multiple sequential scattering, Diffuse Correlation Spectroscopy of tissue primarily measures relative red blood cell motion within vessels},
  journal = {Biomedical Optics Express},
  year = {2011},
  volume = {2},
  number = {7},
  pages = {2047--2054},
  doi = {10.1364/BOE.2.002047},
  pmid = {21750779}
}

@article{Carp2023,
  author = {Carp, Stefan A. and Robinson, Mitchell B. and Franceschini, Maria A.},
  title = {Diffuse correlation spectroscopy: current status and future outlook},
  journal = {Neurophotonics},
  year = {2023},
  volume = {10},
  number = {1},
  pages = {013509},
  doi = {10.1117/1.NPh.10.1.013509},
  pmid = {36704720}
}

@article{BoasDunn2010,
  author = {Boas, David A. and Dunn, Andrew K.},
  title = {Laser speckle contrast imaging in biomedical optics},
  journal = {Journal of Biomedical Optics},
  year = {2010},
  volume = {15},
  number = {1},
  pages = {011109},
  doi = {10.1117/1.3285504},
  pmid = {20210435}
}

@article{Spaide2016,
  author = {Spaide, Richard F.},
  title = {Choriocapillaris Flow Features Follow a Power Law Distribution: Implications for Characterization and Mechanisms of Disease Progression},
  journal = {American Journal of Ophthalmology},
  year = {2016},
  volume = {170},
  pages = {58--67},
  doi = {10.1016/j.ajo.2016.07.023},
  pmid = {27496785}
}

@article{Corvi2021,
  author = {Corvi, Federico and Tiosano, Liran and Corradetti, Giulia and Nittala, Muneeswar Gupta and Lindenberg, Sophiana and Alagorie, Ahmed Roshdy and McLaughlin, John Adam and Lee, Thomas K. and Sadda, Srinivas R.},
  title = {Choriocapillaris flow deficits as a risk factor for progression of age-related macular degeneration},
  journal = {Retina},
  year = {2021},
  volume = {41},
  number = {4},
  pages = {686--693},
  doi = {10.1097/IAE.0000000000002990},
  pmid = {33009219}
}

@article{Choi2012,
  author = {Choi, WooJhon and Baumann, Bernhard and Liu, Jonathan J. and Clermont, Allen C. and Feener, Edward P. and Duker, Jay S. and Fujimoto, James G.},
  title = {Measurement of pulsatile total blood flow in the human and rat retina with ultrahigh speed spectral/Fourier domain OCT},
  journal = {Biomedical Optics Express},
  year = {2012},
  volume = {3},
  number = {5},
  pages = {1047--1061},
  doi = {10.1364/BOE.3.001047},
  pmid = {22567595}
}

@article{Enomoto2021,
  author = {Enomoto, Nobuko and Anraku, Ayako and Tomita, Goji and Iwase, Aiko and Sato, Takashi and Shoji, Nobuyuki and Shiba, Tomoaki and Nakazawa, Toru and Sugiyama, Kazuhisa and Nitta, Koji and Araie, Makoto},
  title = {Characterization of laser speckle flowgraphy pulse waveform parameters for the evaluation of the optic nerve head and retinal circulation},
  journal = {Scientific Reports},
  year = {2021},
  volume = {11},
  number = {1},
  pages = {6847},
  doi = {10.1038/s41598-021-86280-5},
  pmid = {33767305}
}

@article{Puyo:25,
  author = {Puyo, L{\'e}o and Franke, Jonas and Kutzner, Lisa and Pf{\"a}ffle, Clara and Spahr, Hendrik and H{\"u}ttmann, Gereon},
  title = {Measuring choriocapillaris blood flow with laser Doppler optical coherence tomography},
  journal = {Optics Letters},
  year = {2025},
  volume = {50},
  number = {8},
  pages = {2486--2489},
  doi = {10.1364/OL.551061},
  pmid = {40232420}
}

@article{RivaPetrig1995,
  author = {Riva, Charles E. and Petrig, Benno L.},
  title = {Choroidal blood flow by laser Doppler flowmetry},
  journal = {Optical Engineering},
  year = {1995},
  volume = {34},
  number = {3},
  pages = {746--752},
  doi = {10.1117/12.195229}
}

@article{Puyo2019Choroidal,
  author = {Puyo, L{\'e}o and Paques, Michel and Fink, Mathias and Sahel, Jos{\'e}-Alain and Atlan, Michael},
  title = {Choroidal vasculature imaging with laser Doppler holography},
  journal = {Biomedical Optics Express},
  year = {2019},
  volume = {10},
  number = {2},
  pages = {995--1012},
  doi = {10.1364/BOE.10.000995},
  pmid = {30800528}
}

@article{Giammanco2026,
  author = {Giammanco, Manfredi Marco and Bazan Russo, Ginevra Genziana and Giammanco, Marco},
  title = {Laser doppler holography for choroidal blood flow assessment: a systematic review of technical capabilities, validation studies, and clinical applications},
  journal = {Frontiers in Ophthalmology},
  year = {2026},
  volume = {6},
  pages = {1763171},
  doi = {10.3389/fopht.2026.1763171},
  pmid = {42147451}
}

@article{Yazdanfar2003,
  author = {Yazdanfar, Siavash and Rollins, Andrew M. and Izatt, Joseph A.},
  title = {In vivo imaging of human retinal flow dynamics by color Doppler optical coherence tomography},
  journal = {Archives of Ophthalmology},
  year = {2003},
  volume = {121},
  number = {2},
  pages = {235--239},
  doi = {10.1001/archopht.121.2.235}
}

@article{Urs2016,
  author = {Urs, Raksha and Ketterling, Jeffrey A. and Silverman, Ronald H.},
  title = {Ultrafast ultrasound imaging of ocular anatomy and blood flow},
  journal = {Investigative Ophthalmology \& Visual Science},
  year = {2016},
  volume = {57},
  number = {8},
  pages = {3810--3816},
  doi = {10.1167/iovs.16-19538}
}

@article{Henkind1967RPC,
  author  = {Henkind, Philip},
  title   = {Radial Peripapillary Capillaries of the Retina.
             I. Anatomy: Human and Comparative},
  journal = {British Journal of Ophthalmology},
  year    = {1967},
  volume  = {51},
  number  = {2},
  pages   = {115--123},
  doi     = {10.1136/bjo.51.2.115},
  url     = {https://doi.org/10.1136/bjo.51.2.115},
  note    = {PMID: 4959937; PMCID: PMC506341}
}

@article{McLeod1994Choroid,
  author  = {McLeod, D. Scott and Lutty, Gerard A.},
  title   = {High-Resolution Histologic Analysis of the Human
             Choroidal Vasculature},
  journal = {Investigative Ophthalmology \& Visual Science},
  year    = {1994},
  volume  = {35},
  number  = {11},
  pages   = {3799--3811},
  note    = {PMID: 7928177}
}

@article{Yu2015RPC,
  author  = {Yu, Paula K. and Balaratnasingam, Chandrakumar
             and Xu, Jing and Morgan, William H. and Mammo, Zaid
             and Han, Sherry and Mackenzie, Paul and Merkur, Andrew
             and Kirker, Andrew and Albiani, David and Sarunic, Marinko V.
             and Yu, Dao-Yi},
  title   = {Label-Free Density Measurements of Radial Peripapillary
             Capillaries in the Human Retina},
  journal = {PLOS ONE},
  year    = {2015},
  volume  = {10},
  number  = {8},
  pages   = {e0135151},
  doi     = {10.1371/journal.pone.0135151},
  url     = {https://doi.org/10.1371/journal.pone.0135151},
  note    = {PMID: 26252395; PMCID: PMC4529151}
}

@article{Campbell2017RetinalPlexuses,
  author  = {Campbell, J. Peter and Zhang, Miao and Hwang, Thomas S.
             and Bailey, Steven T. and Wilson, David J.
             and Jia, Yali and Huang, David},
  title   = {Detailed Vascular Anatomy of the Human Retina by
             Projection-Resolved Optical Coherence Tomography
             Angiography},
  journal = {Scientific Reports},
  year    = {2017},
  volume  = {7},
  pages   = {42201},
  doi     = {10.1038/srep42201},
  url     = {https://doi.org/10.1038/srep42201},
  note    = {PMID: 28186181}
}

@article{Jia2014GlaucomaOCTA,
  author  = {Jia, Yali and Wei, Eric and Wang, Xiaogang
             and Zhang, Xinbo and Morrison, John C. and Parikh, Mansi
             and Lombardi, Lori H. and Gattey, Devin M.
             and Armour, Rebecca L. and Edmunds, Beth
             and Kraus, Martin F. and Fujimoto, James G.
             and Huang, David},
  title   = {Optical Coherence Tomography Angiography of Optic Disc
             Perfusion in Glaucoma},
  journal = {Ophthalmology},
  year    = {2014},
  volume  = {121},
  number  = {7},
  pages   = {1322--1332},
  doi     = {10.1016/j.ophtha.2014.01.021},
  url     = {https://doi.org/10.1016/j.ophtha.2014.01.021},
  note    = {PMID: 24629312; PMCID: PMC4082728}
}

@article{Moghimi2018GlaucomaVD,
  author  = {Moghimi, Sasan and Zangwill, Linda M.
             and Penteado, Rafaella C. and Hasenstab, Kyle
             and Ghahari, Elham and Hou, Huiyuan
             and Christopher, Mark and Yarmohammadi, Adeleh
             and Manalastas, Patricia Isabel C. and Shoji, Takuhei
             and Bowd, Christopher and Weinreb, Robert N.},
  title   = {Macular and Optic Nerve Head Vessel Density and
             Progressive Retinal Nerve Fiber Layer Loss in Glaucoma},
  journal = {Ophthalmology},
  year    = {2018},
  volume  = {125},
  number  = {11},
  pages   = {1720--1728},
  doi     = {10.1016/j.ophtha.2018.05.006},
  url     = {https://doi.org/10.1016/j.ophtha.2018.05.006},
  note    = {PMID: 29907322}
}

@article{Tansuebchueasai2024ONH,
  author  = {Tansuebchueasai, Natchada and Nishida, Takashi
             and Moghimi, Sasan and Wu, Jo-Hsuan
             and Mahmoudinezhad, Golnoush and Gunasegaran, Gopikasree
             and Kamalipour, Alireza and Zangwill, Linda M.
             and Weinreb, Robert N.},
  title   = {Rate of Initial Optic Nerve Head Capillary Density Loss
             and Risk of Visual Field Progression},
  journal = {JAMA Ophthalmology},
  year    = {2024},
  volume  = {142},
  number  = {6},
  pages   = {530--537},
  doi     = {10.1001/jamaophthalmol.2024.0906},
  url     = {https://doi.org/10.1001/jamaophthalmol.2024.0906},
  note    = {PMID: 38696186; PMCID: PMC11066764}
}

@article{Mandecka2007Flicker,
  author  = {Mandecka, Aleksandra and Dawczynski, Jens
             and Blum, Mathias and M{\"u}ller, Nadja
             and Kloos, Christian and Wolf, Gerhard
             and Vilser, Walthard and Hoyer, Holger
             and M{\"u}ller, Ulrich A.},
  title   = {Influence of Flickering Light on the Retinal Vessels
             in Diabetic Patients},
  journal = {Diabetes Care},
  year    = {2007},
  volume  = {30},
  number  = {12},
  pages   = {3048--3052},
  doi     = {10.2337/dc07-0927},
  url     = {https://doi.org/10.2337/dc07-0927},
  note    = {PMID: 17728481}
}

@article{Sun2019DRProgression,
  author  = {Sun, Zihan and Tang, Fangyao and Wong, Raymond
             and Lok, Jerry and Szeto, Simon K. H.
             and Chan, Jason C. K. and Chan, Carmen K. M.
             and Tham, Clement C. and Ng, Danny S.
             and Cheung, Carol Y.},
  title   = {{OCT} Angiography Metrics Predict Progression of
             Diabetic Retinopathy and Development of Diabetic
             Macular Edema: A Prospective Study},
  journal = {Ophthalmology},
  year    = {2019},
  volume  = {126},
  number  = {12},
  pages   = {1675--1684},
  doi     = {10.1016/j.ophtha.2019.06.016},
  url     = {https://doi.org/10.1016/j.ophtha.2019.06.016},
  note    = {PMID: 31358386}
}

@article{Guo2023Parapapillary,
  author  = {Guo, Xiao and Chen, Yanping and Bulloch, Gabriella
             and Xiong, Kun and Chen, Yifan and Li, Yuting
             and Liao, Huan and Huang, Wenyong and Zhu, Zhuoting
             and Wang, Wei and {Guangzhou Diabetic Eye Study Group}},
  title   = {Parapapillary Choroidal Microvasculature Predicts
             Diabetic Retinopathy Progression and Diabetic Macular
             Edema Development: A Three-Year Prospective Study},
  journal = {American Journal of Ophthalmology},
  year    = {2023},
  volume  = {245},
  pages   = {164--173},
  doi     = {10.1016/j.ajo.2022.07.008},
  url     = {https://doi.org/10.1016/j.ajo.2022.07.008},
  note    = {PMID: 35863493}
}

@article{Yang2026DRPrediction,
  author  = {Yang, Qianhui and Teo, Kelvin Y. C.
             and Hong, Yueheng and Tan, Bingyao and Tan, Rose
             and Schmetterer, Leopold and Mun, Chan Choi
             and Cheung, Gemmy Chui Ming and Wong, Tien Yin
             and Tan, Gavin Siew Wei},
  title   = {Using {OCT} Angiography to Predict Diabetic Retinopathy
             Progression and Vision Decline in a Multiethnic Cohort},
  journal = {Ophthalmology Science},
  year    = {2026},
  volume  = {6},
  number  = {5},
  pages   = {101111},
  doi     = {10.1016/j.xops.2026.101111},
  url     = {https://doi.org/10.1016/j.xops.2026.101111},
  note    = {PMID: 41958713; PMCID: PMC13059305}
}

@article{Zhuang2026DCP,
  author  = {Zhuang, Kallista and Kakihara, Shinji
             and Busza, Anna and Duffy, Brandon
             and Abdelsalam, Mohamed and Fawzi, Amani A.},
  title   = {Baseline Deep Capillary Plexus Ischemia Predicts Earlier
             Diabetic Retinopathy Complications Than Foveal Avascular
             Enlargement},
  journal = {American Journal of Ophthalmology},
  year    = {2026},
  volume  = {292},
  pages   = {322--334},
  doi     = {10.1016/j.ajo.2026.08.014},
  url     = {https://doi.org/10.1016/j.ajo.2026.08.014},
  note    = {PMID: 42600886}
}

@article{Kar2023ChoriocapillarisRMDA,
  author  = {Kar, Deepayan and Corradetti, Giulia
             and Swain, Thomas A. and Clark, Mark E.
             and McGwin, Gerald, Jr. and Owsley, Cynthia
             and Sadda, SriniVas R. and Curcio, Christine A.},
  title   = {Choriocapillaris Impairment Is Associated With Delayed
             Rod-Mediated Dark Adaptation in Age-Related Macular
             Degeneration},
  journal = {Investigative Ophthalmology \& Visual Science},
  year    = {2023},
  volume  = {64},
  number  = {12},
  pages   = {41},
  doi     = {10.1167/iovs.64.12.41},
  url     = {https://doi.org/10.1167/iovs.64.12.41},
  note    = {PMID: 37768273; PMCID: PMC10540875}
}

@article{Pu2026ALSTAR2,
  author  = {Pu, Jiaxin and Gao, Liyan and Thomas, Tracy N.
             and Goerdt, Lukas and Corradetti, Giulia
             and Kar, Deepayan and McGwin, Gerald, Jr.
             and Crosson, Jason N. and Owsley, Cynthia
             and Curcio, Christine A.},
  title   = {Choriocapillaris Flow Signal Deficits Are Associated
             With {AMD} Onset and Early Progression Over Three Years:
             {ALSTAR2} Follow-Up},
  journal = {Investigative Ophthalmology \& Visual Science},
  year    = {2026},
  volume  = {67},
  number  = {3},
  pages   = {51},
  doi     = {10.1167/iovs.67.3.51},
  url     = {https://doi.org/10.1167/iovs.67.3.51},
  note    = {PMID: 41879246; PMCID: PMC13035000}
}

@misc{EyeFlow2026Gap,
  author       = {{DigitalHolography contributors}},
  title        = {{EyeFlow}: open-source retinal low-frequency Doppler spectral analysis},
  year         = {2026},
  howpublished = {Software, version 1.24.0},
  url          = {https://github.com/DigitalHolography/EyeFlow/releases/tag/v1.24.0},
  note         = {Released 23 October 2026; see the low-frequency spectra pipeline; accessed 31 October 2026}
}

@article{Arnould2018EYEMI,
  author  = {Arnould, Louis and Guenancia, Charles and Azemar, Arthur and others},
  title   = {The {EYE-MI} Pilot Study: A Prospective Acute Coronary Syndrome Cohort Evaluated With Retinal Optical Coherence Tomography Angiography},
  journal = {Investigative Ophthalmology \& Visual Science},
  year    = {2018},
  volume  = {59},
  number  = {10},
  pages   = {4299--4306},
  doi     = {10.1167/iovs.18-24090},
  url     = {https://doi.org/10.1167/iovs.18-24090}
}

@article{Sun2021CardiometabolicOCTA,
  author  = {Sun, Michelle T. and Huang, Sonia and Chan, WengOnn and others},
  title   = {Impact of Cardiometabolic Factors on Retinal Vasculature: A {3} $\times$ {3}, {6} $\times$ {6} and {8} $\times$ {8}-mm Optical Coherence Tomography Angiography Study},
  journal = {Clinical \& Experimental Ophthalmology},
  year    = {2021},
  volume  = {49},
  number  = {3},
  pages   = {260--269},
  doi     = {10.1111/ceo.13913},
  url     = {https://doi.org/10.1111/ceo.13913}
}

@article{Yang2025CoronaryOCTA,
  author  = {Yang, Jee Myung and Yang, Dong Hyun and Lee, Seung-Whan and Kwak, Jiehoon and Lee, Yunhan and Kim, Yoon Jeon and Lee, Joo Yong and Sung, Kyung Rim and Yoon, Young Hee},
  title   = {Subclinical Coronary Atherosclerosis and Retinal Optical Coherence Tomography Angiography},
  journal = {JAMA Cardiology},
  year    = {2025},
  volume  = {10},
  number  = {11},
  pages   = {1100--1111},
  doi     = {10.1001/jamacardio.2025.3036},
  url     = {https://doi.org/10.1001/jamacardio.2025.3036}
}

@article{Theuerle2021RetinalFunctionMACE,
  author  = {Theuerle, James D. and Al-Fiadh, Ali H. and Islam, Fakir M. Amirul and Patel, Sheila K. and Burrell, Louise M. and Wong, Tien Yin and Farouque, Omar},
  title   = {Impaired Retinal Microvascular Function Predicts Long-Term Adverse Events in Patients With Cardiovascular Disease},
  journal = {Cardiovascular Research},
  year    = {2021},
  volume  = {117},
  number  = {8},
  pages   = {1949--1957},
  doi     = {10.1093/cvr/cvaa245},
  url     = {https://doi.org/10.1093/cvr/cvaa245}
}

@misc{Bratasz2022Diffuse,
  author        = {Bratasz, Zofia and Martinache, Olivier and Blazy, Yohan and Denis, Ang{\`e}le and Auffret, Coline and Huignard, Jean-Pierre and Rossi, Ethan and Chhablani, Jay and Sahel, Jos{\'e}-Alain and Bonnin, Sophie and Hage, Rabih and Koskas, Patricia and Gatinel, Damien and Vignal, Catherine and Yavchitz, Am{\'e}lie and Vasseur, Vivien and Tadayoni, Ramin and Ducloux, Claire and Ortoli, Manon and Tordjman, Marvin and Tick, Sarah and Mrejen, Sarah and Paques, Michel and Atlan, Michael},
  title         = {Diffuse {Maxwellian} illumination for safe wide-field retinal {Doppler} holography},
  year          = {2022},
  eprint        = {2212.13347},
  archivePrefix = {arXiv},
  primaryClass  = {physics.optics},
  doi           = {10.48550/arXiv.2212.13347},
  note          = {Version 3, revised May 2026}
}

@misc{Auray2025UltrahighSpeed,
  author        = {Auray, Zacharie and Fischer, Yann and Martinache, Olivier and Atlan, Michael},
  title         = {Ultrahigh-speed digital holography for quantitative {Doppler} imaging of the human retina},
  year          = {2025},
  eprint        = {2505.07823},
  archivePrefix = {arXiv},
  primaryClass  = {physics.ins-det},
  doi           = {10.48550/arXiv.2505.07823}
}

@misc{Dubosc2025Holovibes,
  author        = {Dubosc, Marius and Boy-Arnould, Maxime and Guillou, Jules and Gragnic, Titouan and Courselle, Arthur and Herv{\'e}, Gustave and others},
  title         = {{Holovibes}: real-time ultrahigh-speed digital hologram rendering and short-time analysis},
  year          = {2025},
  eprint        = {2508.03911},
  archivePrefix = {arXiv},
  primaryClass  = {physics.optics},
  doi           = {10.48550/arXiv.2508.03911},
  url           = {https://arxiv.org/abs/2508.03911}
}

@misc{Dubosc2025RetinalSegmentation,
  author        = {Dubosc, Marius and Fischer, Yann and Auray, Zacharie and Boutry, Nicolas and Carlinet, Edwin and Atlan, Michael and Geraud, Thierry},
  title         = {Improving segmentation of retinal arteries and veins using cardiac signal in {Doppler} holograms},
  year          = {2025},
  eprint        = {2511.14654},
  archivePrefix = {arXiv},
  primaryClass  = {cs.CV},
  doi           = {10.48550/arXiv.2511.14654},
  url           = {https://arxiv.org/abs/2511.14654},
  note          = {Version 2, revised February 2026}
}

@misc{Buisson2026WaveformShape,
  author = {Buisson, Gr{\'e}goire and Paquet, Chlo{\'e} and Doucet, Lancelot and Gouyoumdjian, David and Boy Arnould, Maxime and Fischer, Yann and Auray, Zacharie and Atlan, Michael},
  title = {Transportable retinal {Doppler} holography waveform-shape metrics for functional microvascular assessment},
  year = {2026},
  howpublished = {Zenodo},
  doi = {10.5281/zenodo.18895922},
  url = {https://doi.org/10.5281/zenodo.18895922}
}

@article{Srinivasan2010QuantitativeOCT,
  author = {Srinivasan, Vivek J. and Sakad{\v{z}}i{\'c}, Sava and Gorczynska, Iwona and Ruvinskaya, Svetlana and Wu, Weicheng and Fujimoto, James G. and Boas, David A.},
  title = {Quantitative cerebral blood flow with optical coherence tomography},
  journal = {Optics Express},
  year = {2010},
  volume = {18},
  number = {3},
  pages = {2477--2494},
  doi = {10.1364/OE.18.002477},
  url = {https://doi.org/10.1364/OE.18.002477},
  pmid = {20174075}
}

@article{Makita2006Angiography,
  author = {Makita, Shuichi and Hong, Youngjoo and Yamanari, Masahiro and Yatagai, Toyohiko and Yasuno, Yoshiaki},
  title = {Optical coherence angiography},
  journal = {Optics Express},
  year = {2006},
  volume = {14},
  number = {17},
  pages = {7821--7840},
  doi = {10.1364/OE.14.007821},
  url = {https://doi.org/10.1364/OE.14.007821},
  pmid = {19529151}
}

@article{Wang2007OpticalAngiography,
  author = {Wang, Ruikang K. and Jacques, Steven L. and Ma, Zhenhe and Hurst, Sawan and Hanson, Stephen R. and Gruber, Andras},
  title = {Three dimensional optical angiography},
  journal = {Optics Express},
  year = {2007},
  volume = {15},
  number = {7},
  pages = {4083--4097},
  doi = {10.1364/OE.15.004083},
  url = {https://doi.org/10.1364/OE.15.004083},
  pmid = {19532651}
}

@article{Migacz2019MHzOCTA,
  author = {Migacz, Justin V. and Gorczynska, Iwona and Azimipour, Mehdi and Jonnal, Ravi and Zawadzki, Robert J. and Werner, John S.},
  title = {Megahertz-rate optical coherence tomography angiography improves the contrast of the choriocapillaris and choroid in human retinal imaging},
  journal = {Biomedical Optics Express},
  year = {2019},
  volume = {10},
  number = {1},
  pages = {50--65},
  doi = {10.1364/BOE.10.000050},
  url = {https://doi.org/10.1364/BOE.10.000050},
  pmid = {30775082}
}

@article{Zhang2018CCFlowVoids,
  author = {Zhang, Qinqin and Zheng, Fang and Motulsky, Elie H. and Gregori, Giovanni and Chu, Zhongdi and Chen, Chieh-Li and Li, Chunxia and de Sisternes, Luis and Durbin, Mary and Rosenfeld, Philip J. and Wang, Ruikang K.},
  title = {A Novel Strategy for Quantifying Choriocapillaris Flow Voids Using Swept-Source {OCT} Angiography},
  journal = {Investigative Ophthalmology \& Visual Science},
  year = {2018},
  volume = {59},
  number = {1},
  pages = {203--211},
  doi = {10.1167/iovs.17-22953},
  url = {https://doi.org/10.1167/iovs.17-22953},
  pmid = {29340648}
}

@article{Wei2019HDROCTA,
  author = {Wei, Xiang and Hormel, Tristan T. and Pi, Shaohua and Guo, Yukun and Jian, Yifan and Jia, Yali},
  title = {High dynamic range optical coherence tomography angiography ({HDR-OCTA})},
  journal = {Biomedical Optics Express},
  year = {2019},
  volume = {10},
  number = {7},
  pages = {3560--3571},
  doi = {10.1364/BOE.10.003560},
  url = {https://doi.org/10.1364/BOE.10.003560},
  pmid = {31360605}
}
\end{document}